\documentclass[prd,amsfonts,amsmath,twocolumn,superscriptaddress,nofootinbib]{revtex4}

\usepackage{color}
\usepackage{graphicx} 
\usepackage{epstopdf}%

\usepackage{mathpazo}      

\newcommand {\be} {\begin{equation}} 
\newcommand {\ba}{\begin{eqnarray}} 
\newcommand {\ee} {\end{equation}} 
\newcommand{\ea} {\end{eqnarray}}

\renewcommand{\epsilon}{\varepsilon}

\begin{document}

\title{
High-Multipole Excitations of Hydrogen-Like Atoms by \\ Twisted Photons near Phase Singularity}

\author{Andrei Afanasev}

\affiliation{Department of Physics,
The George Washington University, Washington, DC 20052, USA}

\author{Carl E.\ Carlson}

\affiliation{Department of Physics, The College of William and Mary in Virginia, Williamsburg, VA 23187, USA}

\author{Asmita Mukherjee}

\affiliation{Department of Physics, Indian Institute of Technology Bombay, Powai, 
Mumbai 400076, India}

\date{\today}

\begin{abstract}

We calculate transition amplitudes and cross sections for excitation of hydrogen-like atoms by the twisted photon states, or photon states with angular momentum projection on the direction of propagation 
exceeding $\hbar$. If the target atom is located at distances of the order of atomic size near the phase singularity in the vortex center, the transition rates into the states with orbital angular momentum $l_f>1$ 
become comparable with the rates for electric dipole transitions. It is shown that when the transition rates are normalized to the local photon flux, the resulting 
cross sections for $l_f>1$  are singular near the optical vortex center.  Relation to the ``quantum core'' concept introduced by Berry and Dennis is discussed.

\end{abstract}

\maketitle

\section{Introduction and Motivation}	\label{sec:intro}

Optical vortices, or Orbital Angular Momentum (OAM) beams are a subject of active research since the publication of Allen and collaborators  \cite{Allen:1992zz} .
At a quantum level such beams can be described in terms of "twisted photons" \cite{molina2007nature}. Comprehensive reviews of the subject are available in the literature \cite{Yao11,Padgett2015}. 

The main focus of this study is interaction of the twisted photons with individual atoms or ions. In previous work we developed a formalism for calculating photoexcitation of hydrogen-like atoms by the twisted photons and demonstrated novel quantum selection rules \cite{Afanasev:2013kaa}, with recoil effects considered in Ref.\cite{Afanasev2014}. In particular, it was shown in Ref. \cite{Afanasev:2013kaa} that atomic transitions with 
$\Delta l$=3 caused by the twisted photons have nonzero transition amplitude at the zero-intensity center of Bessel beam. Another step relevant to the present study was made in Ref.\cite{Scholz2014} that demonstrated, in particular, that the transition amplitudes with twisted photons can be presented in a simple factorized form in terms of plane-wave photon transitions. Absorption of the twisted light by many-electron atoms and ions was considered in Ref.\cite{Surzhikov15}.
Recent  theoretical work in this field includes excitation of Rydberg atoms with OAM beams \cite{Rodrigues2015}, additional consideration of quantum selection rules with recoil effects \cite{Jaregui2015}, theoretical analysis of angular momentum transfer to atomic electrons \cite{Mondal2014}, and optical vortex interaction with multi-electron atoms formulated in the impact-parameter space \cite{Kaplan15}.

An outstanding question is an ability of twisted light to pass its angular momentum to the internal quantum states of the atomic (or molecular) target. It has been pointed out previously \cite{Babiker2002} that it is not possible for electric-dipole (E1) excitations that pass extra angular momentum only to center-of-mass motion of the entire atom. Work by Picon {\it et al.} \cite{Picon10} demonstrated that during photoionization of atoms, the knocked-out electrons carry angular momenta that reproduce the angular momentum of the incoming photons.
In a new and significant experimental development, the authors of  Ref.\cite{Schmiegelow2015}, using an ion trap, analyzed excitation of a single $^{40}$Ca$^+$ ion with a vortex laser beam, demonstrated nonzero rate of electric quadrupole E2 transition in the zero-intensity center of the optical vortex, and confirmed that the optical angular momentum may be passed to the internal degrees of freedom of an atom. For E2 transitions this remarkable "excitation in the dark" effect is due to nonzero field gradients, as was shown in Ref.\cite{Schmiegelow2012}. 

In the present article we analyze quadrupole E2 ($\Delta l=2$) and octupole E3 ($\Delta l=3$) transitions caused by the twisted photons, show that "excitation in the dark" takes place for all higher multipoles, and compare the atomic excitation rates and cross sections with their plane-wave analogues as a function of atomic target position within the optical vortex.
Sections II and III of this paper review the formalism for twisted photons and for calculating atomic photoexcitation with the twisted photons, respectively.  Sec. IV discusses angular momentum selection rules for photoexcitation, emphasizing possibilities for twisted photons that are impossible for plane wave photons.  Sec. V discusses the photoexcitation cross sections, in particular showing non-zero cross sections for situations where the twisted photon beam has zero field at its center and that center hits the atom directly.  The amplitudes in these cases are proportional to the gradient or curvature of the field profile, and are characterized by jumps of two or three (or more) units of angular momentum projection.  Finally, Sec. VI offers some closing comments.


\section{Definition of Twised-Photon States}			\label{sec:one}


We define the twisted-photon states according to~\cite{Jentschura:2010ap,Jentschura:2011ih}, that can be viewed as extensions of the nondiffractive Bessel modes described in~\cite{Durnin:1987,Durnin:1987zz}. More detail is given in~\cite{Afanasev:2013kaa}.   These states correspond to superposition of TE and TM Bessel modes introduced in Ref. \cite{Jauregui:2004}; see also Appendix of Ref.\cite{Afanasev2014}.

A twisted photon state with symmetry axis passing through the origin, can be given as a superposition of plane waves and in Hilbert space can be written as, 
\begin{align}
\label{eq:twisteddefinition}
| \kappa m_\gamma k_z \Lambda \rangle 
&= \sqrt{\frac{\kappa}{2\pi}} \  \int \frac{d\phi_k}{2\pi} (-i)^{m_\gamma} e^{im_\gamma\phi_k}  \,
	|\vec k, \Lambda\rangle		\,.
\end{align}
The component states on the right are plane wave states, all with the same longitudinal momentum $k_z$, the same transverse momentum magnitude $\kappa = |\vec k_\perp|$, and the same plane wave helicity $\Lambda$ (in the directions $\vec k$).  The angle  $\phi_k$  is the azimuthal angle of vector $\vec k$, and with the phasing shown, $m_\gamma$ is the total angular momentum in the $z$ direction, with the possibility that $m_\gamma \gg 1$.  We also define a pitch angle   $\theta_k = \arctan (\kappa/k_z)$, and $\omega = | \vec k |$. The phase singularity of this beam is located on the beam symmetry axis.

The electromagnetic potential of the twisted photon in coordinate space is 
\begin{align}
\label{eq:twistedwave}
\mathcal A^\mu_{\kappa m_\gamma k_z \Lambda}(t,\vec r)
&= \sqrt{\frac{\kappa}{2\pi}} \, e^{-i\omega t} \nonumber\\
&\times	\int \frac{d\phi_k}{2\pi} (-i)^{m_\gamma} e^{im_\gamma\phi_k}  \,
	\epsilon^\mu_{\vec k,\Lambda} e^{i \vec k {\cdot} \vec r_1}	.
\end{align}
The polarization vectors are~\cite{Jentschura:2010ap,Jentschura:2011ih,Afanasev:2013kaa}
\be
\label{eq:epsilonexpand}
\epsilon^\mu_{\vec k \Lambda} \!\! = \!
	e^{-i\Lambda\phi_k} \! \cos^2\frac{\theta_k}{2} \eta^\mu_\Lambda
	+ e^{i\Lambda\phi_k} \! \sin^2\frac{\theta_k}{2} \eta^\mu_{-\Lambda}
	+ \frac{\Lambda}{\sqrt{2}} \sin\theta_k  \eta^\mu_0
\ee
with $4$-dimensional unit vectors,
\be
\eta^\mu_{\pm 1} = \frac{1}{\sqrt{2}}  \left( 0,\mp 1,-i,0 \right)	\,,
\quad \eta^\mu_0 =  \left( 0,0,0,1 \right)	\,.
\ee
The electromagnetic potential then has a form
\begin{align}
\label{eq:twistedwf}
\mathcal A^\mu_{\kappa m_\gamma k_z \Lambda}(x) &= e^{-i(\omega t - k_z z)}	
\sqrt{\frac{\kappa}{2\pi}}  \, \nonumber\\ &	\Bigg\{
	\frac{\Lambda}{\sqrt{2}} e^{im_\gamma\phi_\rho} \sin\theta_k
	J_{m_\gamma}(\kappa\rho) \, \eta^\mu_0			\nonumber\\[1ex]
& \quad + i^{-\Lambda} e^{i(m_\gamma-\Lambda)\phi_\rho}  \cos^2\frac{\theta_k}{2} 
	J_{m_\gamma-\Lambda}(\kappa\rho) \, \eta^\mu_\Lambda	\nonumber\\[1ex]
& \quad + i^{\Lambda}  e^{i(m_\gamma+\Lambda)\phi_\rho}  \sin^2\frac{\theta_k}{2} 
	J_{m_\gamma+\Lambda}(\kappa\rho) \, \eta^\mu_{-\Lambda}
	\Bigg\}	\,.
\end{align}

The corresponding magnetic field is 
\begin{align}
\label{eq:magfield}
B_\rho &=  i\omega \Lambda \sqrt{\frac{\kappa}{4\pi}} e^{i(k_z z -\omega t + m_\gamma\phi)}
	\nonumber\\
& \quad \times \left( \sin^2\frac{\theta_k}{2} J_{m_\gamma+\Lambda}(\kappa\rho) 
		+ \cos^2\frac{\theta_k}{2} J_{m_\gamma-\Lambda}(\kappa\rho) \right) \,,	\nonumber\\
B_\phi &=  \omega \Lambda \sqrt{\frac{\kappa}{4\pi}} e^{i(k_z z -\omega t + m_\gamma\phi)}
	\nonumber\\
& \quad \times \left( \sin^2\frac{\theta_k}{2} J_{m_\gamma+\Lambda}(\kappa\rho) 
		- \cos^2\frac{\theta_k}{2} J_{m_\gamma-\Lambda}(\kappa\rho) \right) \,,	\nonumber\\
B_z &= \omega \Lambda \sqrt{\frac{\kappa}{4\pi}} e^{i(k_z z -\omega t + m_\gamma\phi)}
	\sin\theta_k J_{m_\gamma}(\kappa\rho)	\,,
\end{align}
and the electric field is just 90$^\circ$ out of phase with the magnetic field, $\vec E = i \vec B$.  

The local energy flux (needed for evaluation of the cross sections) is given by (c.f. Eq.(27) of Ref.\cite{BliokhAlonso10})
\begin{align}
\label{eq:flux}
& f(\rho)  = \cos(\theta_k) (|E|^2+|B|^2)/4= 
 \cos(\theta_k) \frac{\kappa\omega^2 }{2\pi} \\
& \left( \cos^4\frac{\theta_k}{2} J^2_{m_\gamma-\Lambda}(\kappa\rho)+\sin^4\frac{\theta_k}{2} J^2_{m_\gamma+
\Lambda}(\kappa\rho)+ \frac{\sin^2\theta_k}{2}J^2_{m_\gamma}(\kappa\rho) \right). \nonumber \\ \nonumber
\end{align}
The use of the above canonical-momentum expression is essential, since Poynting vector alone does not represent the full energy flux of a twisted photon beam \cite{BliokhAlonso10,Bliokh2015}.


\section{Atomic photoexcitation with twisted photons}			


Here we briefly review the formalism of atomic photoexcitation by the twisted photons worked out previously by the authors \cite{Afanasev:2013kaa,Afanasev2014} and relate our formulae for the transition amplitudes to the more recent results of \cite{Scholz2014}. 

Consider the excitation by a twisted photon of a hydrogen-like atom from the ground state. The photon's wave front travels in the $z$-direction and the axis of the twisted photon is displaced from the nucleus of the atomic target by some distance in the $x$-$y$ plane which we will call an impact parameter $\vec b$.  The transition matrix element is
\begin{align}
S_{fi} &= -i \int dt  \langle n_f l_f m_f | H_1 
	| n_i l_i m_i; \kappa m k_z \Lambda \rangle				\,,
\end{align}
where the non-relativistic interaction Hamiltonian is given by
\be
\label{eq:hamiltonian}
H_1 = - \frac{e}{m_e} \vec A \cdot \vec p \,,
\ee
and we use standard notation $(n,l,m)$ for the quantum numbers of initial and final states of a hydrogen atom.

\begin{figure}[b]
\begin{center}
\includegraphics{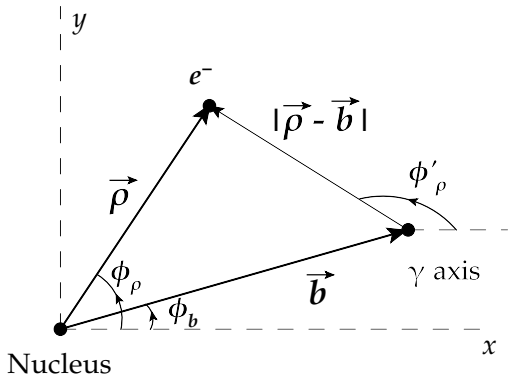}
\caption{Relative positions of atomic state and photon axis, as projected onto the $x$-$y$ plane, with the origin at the nucleus of the atom.}
\label{offaxiscoord}
\end{center}
\end{figure}

\begin{align}
\label{eq:genresult}
S_{fi} &= -2\pi \delta(E_f-E_i-\omega)  \frac{e}{m_e a_0}  \,  
	\sqrt{\frac{2\pi\kappa}{3}}	\,	i^{-\Lambda} e^{i(m_\gamma-m_f)\phi_b} \, 
					\nonumber\\
&\times	 J_{m_f-m_\gamma}(\kappa b)\Bigg\{
		 \cos^2\frac{\theta_k}{2} \, g_{n_f l_f m_f \Lambda}	+ 		
		 \frac{i}{\sqrt{2}} \sin\theta_k 	\, g_{n_f l_f m_f 0}\nonumber\\	&- \sin^2\frac{\theta_k}{2} \,  g_{n_f l_f m_f, -\Lambda}
	\Bigg\}		\nonumber\\
&\stackrel{\rm def}{=} 2\pi \delta(E_f-E_i-\omega)  \, \mathcal M_{n_f l_f m_f \Lambda}(b)
	\,.
\end{align}

\noindent  The dimensionless atomic factors are
\begin{align}
\label{eq:reduced}
&g_{n_f l_f m_f \Lambda} \nonumber \\ &\equiv  - a_0
	\int_0^\infty r^2 dr \ R_{n_f l_f}(r)  \, R'_{10}(r)  \nonumber \\
	& \times \int_{-1}^1 d(\cos\theta_r) \, J_{m_f-\Lambda}(\kappa \rho) \, 
	Y_{l_f m_f}(\theta_r,0) \, Y_{1 \lambda}(\theta_r,0) e^{i k_z z}		\,,
\end{align} 
where $a_0$ is the Bohr radius, and $R_{n_f l_f}(r)$ are radial wave functions of the excited atomic state.

In agreement with Ref.\cite{Scholz2014} (c.f. Eqs.(19,20)), we can verify that the above amplitude from Eq.(\ref{eq:genresult}) is proportional to the plane-wave amplitude $\mathcal M^{\rm (pw)}$ times $d$-functions that only depend on the pitch angle $\theta_k$:
\begin{align}
\Bigg\{
		 \cos^2\frac{\theta_k}{2} \, g_{n_f l_f m_f \Lambda}	+ 		
		 \frac{i}{\sqrt{2}} \sin\theta_k 	\, g_{n_f l_f m_f 0}\nonumber\\	- \sin^2\frac{\theta_k}{2} \,  g_{n_f l_f m_f, -\Lambda}
	\Bigg\} \propto \nonumber \\ d^{l_f}_{m_f\Lambda}(\theta_k) \mathcal M^{\rm (pw)}_{n_f l_f \Lambda \Lambda}(\theta_k=0)
\end{align}
The factorized form of the transition amplitude is convenient for comparison of twisted-photon vs plane-wave photo absorption:
\begin{align}
\label{eq:factorized}
&|{\cal M}_{n_f l_f m_f \Lambda}(b)|   =  \nonumber \\  &\left |\sqrt{\frac{\kappa}{2\pi}}J_{m_f-m_\gamma}(\kappa b)d^{l_f}_{m_f\Lambda}(\theta_k) \mathcal M^{\rm (pw)}_{n_f l_f \Lambda \Lambda}(\theta_k=0)\right |
\end{align}
It should be noted that the above form of the twisted-photon absorption amplitude contains the details of atomic structure in a factorized plane-wave amplitude $\cal M^{\rm (pw)}$, while the novel features arising from the phase and spatial structure of the OAM beams are contained in Wigner and Bessel factors independent of the atomic wave functions. In the derivation we only assumed that the atom is much smaller than the photon wavelength, therefore the above result applies to arbitrary targets, as long as this long-wavelength condition is met.

\section{Transition amplitudes near the optical vortex center}

The region of the transverse beam profile near the optical vortex center, where the impact parameter $b\to 0$, corresponds to the phase singularity of the Bessel beam.
Remembering that $\kappa=\omega \sin(\theta_k)$, we find for small values of impact parameter $b$ and the pitch angle $\theta_k$
\be
\label{eq:lowb}
 |{\cal M}_{n_f l_f m_f \Lambda}(b)| \approx \sqrt{\frac{\kappa}{2\pi}}(\theta_k)^{|m_\gamma-\Lambda|}(\omega b)^{|m_\gamma-m_f|}(\omega a_0)^{(l_f-1)}
 \,,
\ee
where the last factor $(\omega a_0)^{(l_f-1)}$ is a familiar result for plane-wave photons.

From Eq.(\ref{eq:lowb}) we can estimate relative strength of excitation by a twisted photon with $m_\gamma>1$, $\Lambda=1$ into atomic states with different OAM $l_f$:

\noindent (a) Electric dipole ($\Delta l=1$) transition $l_f=1$,
\be
\label{eq:edip}
{\cal M}(l_f=1)\propto (\theta_k)^{|m_\gamma-1|}(\omega b)^{(m_\gamma-1)}
\ee

\noindent (b) Transitions with $l_f= m_\gamma$, the largest amplitude for $b\to 0$ corresponding to $m_\gamma=m_f$ reads
\be
\label{eq:lowb-anyl}
{\cal M}(l_f=m_\gamma)\propto (\theta_k)^{|m_\gamma-1|} (\omega a_0)^{(l_f-1)}
\ee
The above equation implies that the transition amplitude into $m_\gamma=m_f$ state is finite at the vortex center even though the field strength is zero there: $|A|\propto (\kappa b)^{(m_\gamma-1)}$ if we expand Bessel function $J_{m_\gamma-\Lambda}(\kappa b)$ near $b\to 0$ in Eq.(\ref{eq:twistedwf}).
Therefore  the quadrupole E2 transition ($\Delta l=2$) in the vortex center is due to the field {\it gradient} $A'$, while the octupole E3 transition ($\Delta l=3$) is due to field's second derivative $A''$ or {\it curvature}.

If $m_\gamma\ge l_f$, the ratio of $\Delta l>1$ to $\Delta l=1$ amplitudes is independent of either the pitch angle or photon energy and scales as
\be
\label{eq:ratamps}
\frac{{\cal M}(m_\gamma\ge l_f>1)}{{\cal M}(l_f=1)}\propto\left(\frac{a_0}{b}\right )^{(l_f-1)}
\ee
while for $m_\gamma<l_f$ the following expression holds
\be
\frac{{\cal M}(l_f=m_\gamma)}{{\cal M}(l_f=1)}\propto \frac{(\omega a_0)^{(l_f-1)}}{(\omega b)^{(m_\gamma-1)}}
\ee

It follows from Eqs.(\ref{eq:edip}--\ref{eq:ratamps}) that the transition rates into $l_f>1$ atomic states are relatively enhanced with the 
application of OAM beams with matching $m_\gamma=l_f$. Electric-dipole transition strength scales with the field strength of Bessel beam as $(\omega b)^{(m_\gamma-1)}$ and turns to zero as $b\to 0$, while $l_f=m_\gamma>1$ amplitude remains finite in the vortex center. However, to make $\Delta l>1$ transitions similar in magnitude to standard 
electric dipole $\Delta l=1$  transition, the atom has to be placed within the distances from the vortex center of the order of atomic 
size $a_0$. For higher values of $b\gg a_0$ the transitions with $\Delta l=1$ remain dominant due to the fact that in the considered optical domain $(\omega a_0)\ll 1$, i.e. the wavelength is much larger than the atomic size.

The discussed behavior of the transition amplitudes has direct implications for the photoexcitation cross sections. Corresponding expressions 
for cross sections have been worked out previously \cite{Afanasev:2013kaa,Scholz2014}. Calculation of the cross sections requires definition of the photon 
flux for normalization. In Refs.\cite{Afanasev:2013kaa,Scholz2014}, the flux integrated over the transverse size of the beam was used, and under 
this convention comparison of cross sections is similar to comparing squares of the transition amplitudes discussed above.

As follows from Eqs.(\ref{eq:factorized},\ref{eq:lowb}) the pitch angle $\theta_k$ is an important parameter to determine the absorption rates of the twisted photons with $m_\gamma>1$ ; dependence on $\theta_k$is a combined effect of Bessel and Wigner functions in the expression (\ref{eq:factorized}). If $\theta_k \to 0$, the Wigner function is only nonzero for $m_f=\Lambda$, enforcing total angular 
momentum  conservation, so that the only surviving amplitude matches quantum selection rules form plane-wave photon absorption. On the other hand, for larger values of $\theta_k$ all the final states with $-l_f\le m_f \le l_f$ become populated (c.f. Fig. 4 of Ref.\cite{Afanasev:2013kaa}), while $b$-dependence of the relative strength of the amplitudes is determined by the Bessel function $J_{m_f-m_\gamma}(\kappa b) $.

\begin{figure}[t]
\begin{center}
\includegraphics[width = 84 mm]{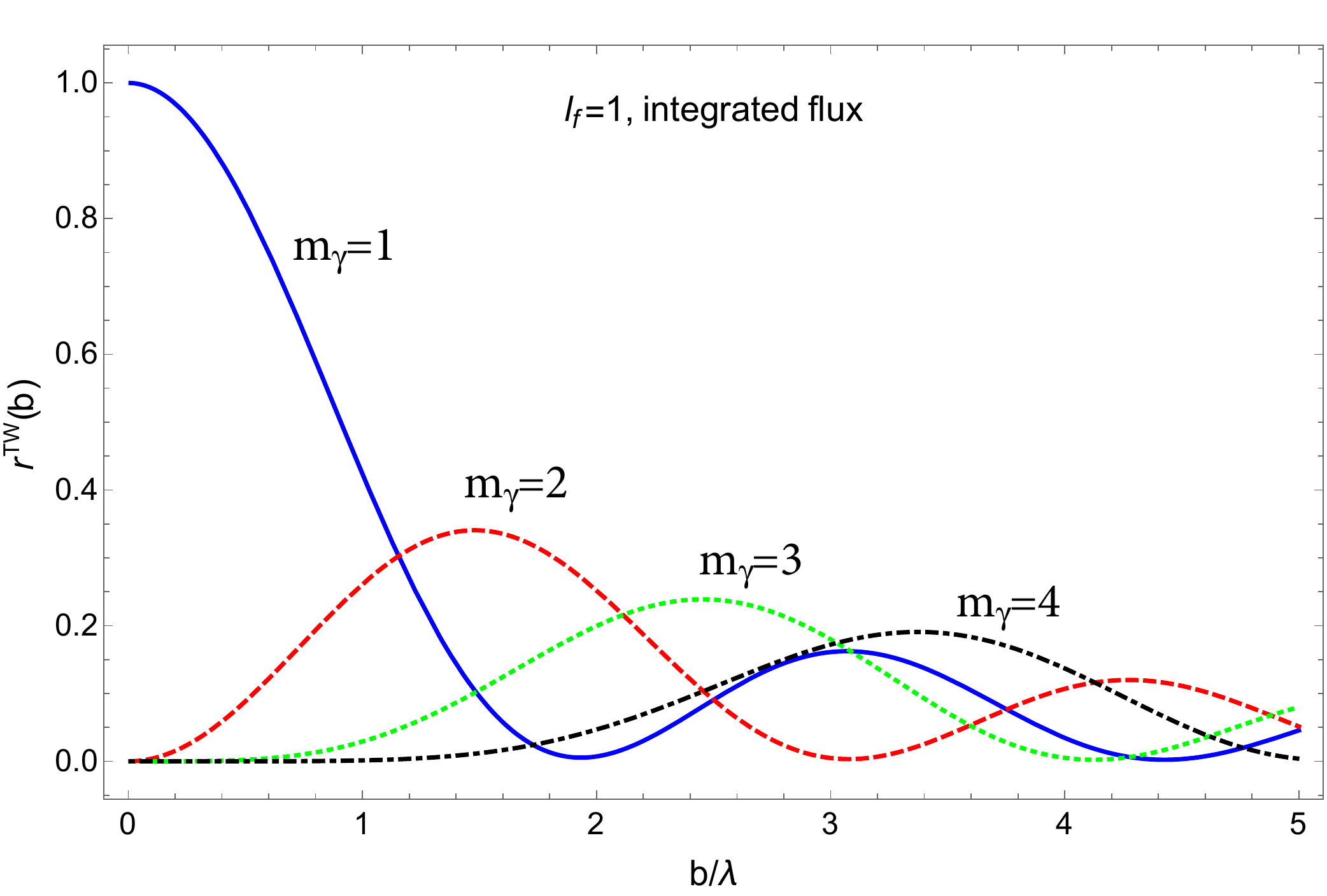}
\caption{ The ratio $r^{TW}(b)$ using an integrated flux $f$  for various values of $m_\gamma$ for excitation of $l_f=1$ state. The curve styles denote the photon's angular momentum projection: $m_\gamma=1$ is the blue solid curve, $m_\gamma=2$ is red and dashed, $m_\gamma=3$ is green and dotted, $m_\gamma=4$ is black and dot-dashed. Since the excitation is caused by electric dipole transition, the curves scale with the local photon flux. }
\label{fig:flux}
\end{center}
\end{figure}

\begin{widetext}

\begin{figure}[t]
\begin{center}
\includegraphics[width = 180 mm]{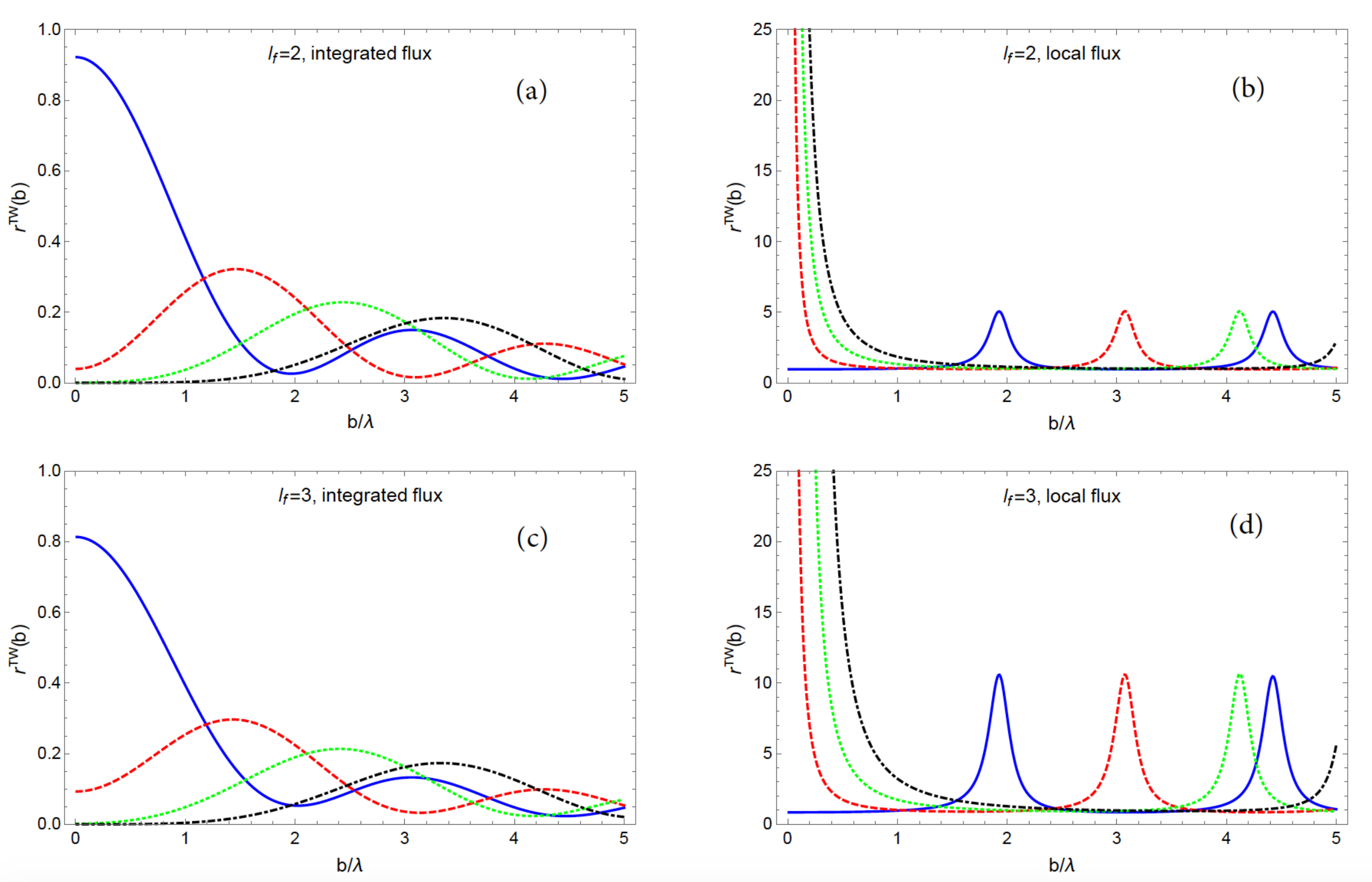}
\end{center}
\caption{(a) The ratio $r^{TW}(b)$ using either integrated (a,c)  or local (b,d) flux $f$  for various values of $m_\gamma$ for excitation of $l_f=2$ state (a,b) and $l_f$=3 (c,d). The curve styles are the same as in Fig.\ref{fig:flux}. Deviation of the curves in (a,c) from same-style curves in Fig.\ref{fig:flux} indicate the deviation from the plane-wave limit, while their ratios yield the plots in (b,d). The plane-wave limit corresponds to $r^{TW}(b)$=1 in (b,d). }
\label{fig:ratios}
\end{figure}

\end{widetext}

\section{Cross Sections of Twisted-Photon Absorption}

We need to adopt a convention to evaluate the reaction cross sections if the photon flux has a variable density across the transverse dimensions. Averaging the photon flux over the transverse beam profile was used in previous work \cite{Afanasev:2013kaa,Scholz2014}. A recent experiment on vortex-beam absorption by single trapped ions localized in the center of optical vortex \cite{Schmiegelow2015} demonstrated feasibility of the measurements a function of the impact parameter $b$. Motivated by the advances in experimental techniques, we evaluate the excitation rate from Eqs.(\ref{eq:genresult},\ref{eq:reduced}) as a function of $b$, and present the results with both integrated flux and unintegrated flux $f(b)$ given by Eq.(\ref{eq:flux}).

\begin{equation}
\sigma^{(m_\gamma)}_{n_f l_f m_f \Lambda} = 2 \pi \delta(E_f -E_i-\omega_\gamma)
 \frac{ |\mathcal M^{(m_\gamma)}_{n_f l_f m_f \Lambda}(b)|^2 }{f} \,.
\end{equation}
where summation over final spins and averaging over initial spins is implied and the photon flux $f$ may be either unintegrated $f(b)$ (given by Eq.(\ref{eq:flux})) or integrated over the transverse beam profile. 

We introduce a quantity $r^{\rm tw}$ that compares twisted-wave and plane-wave cross sections of atomic excitation for different values of the impact parameter $b$. \footnote{In the previous work  \cite{Afanasev:2013kaa} we averaged both the flux and the transition rates over the transverse beam profile and found it  to be close to unity (or, more accurately, $r^{\rm tw}=1.02$ for $\theta_k=0.2$).} 

\begin{equation}
r^{\rm tw}(b) = \frac{ \sum_{m_f = - l_f}^{m_f=l_f}
	\sigma_{n_f l_f m_f \Lambda} } 
	{\sigma_{n_f l_f \Lambda \Lambda}^{(\rm pw)}}	\,.
\end{equation}

If the same average flux is used for normalizing plane-wave and twisted-wave cross sections, we verified numerically that for $l_f$=1 ($E1$-transition) this quantity is proportional to the beam intensity profile $r^{\rm tw}(b)\propto f(b)$, for any values of $b$ or $m_\gamma$, see Fig.\ref{fig:flux}. This result can be anticipated, since for the electric-dipole case the transition amplitude is proportional to the electromagnetic field strength, as can be seen also from Eq.(\ref{eq:factorized}) and also follows from the analytical expressions of  Ref. \cite{Scholz2014}. This observation implies that the vortex-beam interaction via the largest electric-dipole transitions while propagating through the medium is independent of the transverse beam intensity profile. If the local (unintegrated) energy flux $f(b)$ is used for normalizing the cross sections, then $r^{\rm tw}(b)=1$ for $E1$-transitions, in agreement with the conclusions of earlier studies \cite{Babiker2002}.

Next, we consider photoexcitation of $l_f=2$ and 3 states with Bessel beam with different $m_\gamma=1$ to 4. The choice of parameter is as follows: The pitch angle is $\theta_k=0.2$; the impact parameter $b$ is measured in units of wavelength $\lambda$ determined by the excited energy level of hydrogen, in this example the principal quantum number of the excited state $n_f=4$ and $\lambda$=97 nm. The results for $r_{\rm tw}(b)$ with integrated flux are shown in Fig.\ref{fig:ratios}(a,c). Note that the total beam power was the same for all choices of $m_\gamma$, therefore plots in Fig.\ref{fig:ratios}(a,c) represent the rates relative to the plane wave case. Let us discuss the main features of the presented plots. First, we observe an interesting effect for $m_\gamma=1$ Bessel beam, namely, suppression of the transition rate (compared to plane waves of the same local intensity) near the vortex center by about 7 per cent for E2 and almost 20 per cent for E3 transitions, respectively.  We can attribute this suppression to the Wigner function $d^{l_f}_{m_f\Lambda}(\theta_k) $ of Eq.(\ref{eq:factorized}) that acts as a probability amplitude for transitions into $m_f=m_\gamma$ excited states that are the only ones allowed in the vortex center. 

Second, and most remarkably, in Fig.\ref{fig:ratios}(a,c) one can see that the excitation rates are finite in the vortex center at zero field intensity ($b=0$) when $l_f\ge m_\gamma>1$, while the rate is zero when $m_\gamma>l_f$. It follows from the behavior of Bessel function in the origin, $J_{m_f-m_\gamma}(0)=\delta_{m_\gamma m_f}$. This ''excitation in the dark" effect for electric-quadrupole $\Delta l=2$ transitions caused by LG beams on $^{40}$Ca$^+$ ions was predicted in Ref.\cite{Schmiegelow2012} and demonstrated experimentally in Ref.\cite{Schmiegelow2015}. Since the photon wavelength in the experiment \cite{Schmiegelow2015} was $\lambda=729$nm, which is much larger that the size of $^{40}$Ca$^+$ ion, our formalism for comparing with plane-wave photon absorption may be applied for this case, too, if the beam profile is replaced with Laguerre-Gaussian (LG). Note that LG beams have more steep fall-off of intensity away from the vortex center, so we expect that the region near vortex center is more pronounced in LG case. Extension of our formalism to LG beams will be a subject of future work.
In the example shown in  Fig.\ref{fig:ratios}(a,c), the vortex-center rates are nonzero for $m_\gamma$=2, $l_f=2,3$ case and for $m_\gamma=3$, $l_f=3$; for the latter case it is numerically smaller by an order of magnitude due to a factor $|d^{l_f}_{m_f\Lambda}(\theta_k)|_{\theta_k=0.2}^2$ from Eq.(\ref{eq:factorized}). The plots in Fig.\ref{fig:ratios}(b,d) correspond to the division by unintegrated flux $f(b)$. In this case $r^{\rm tw}(b)=1$ for $l_f=1$, while for higher angular momentum of the excited states we observe that the cross section normalized to a local flux is either enhanced for lower-intensity regions or singular for the zero-intensity vortex center.

The observed singular behavior can be understood using our analytical formalism. For $l_f>0$ we can use the small-$b$ expansion (\ref{eq:edip},\ref{eq:lowb-anyl}) to obtain  $b$-dependence at $(\omega b)\ll 1$ for $m_\gamma=l_f$
\be
r^{\rm tw}(b)\to \left(\frac{a_0}{b}\right )^{(2m_\gamma-2)} \\.
\ee

This power-like singularity is independent on the photon wavelength or pitch angle  $\theta_k$, while the Bohr's radius $a_0$ sets the scale for the dependence on impact parameter $b$.

Remembering that $a_0\to 0$ in the classical limit $\hbar\to 0$, we can classify this singular behavior as a quantum effect that appears for the transition rates normalized to the local intensity of light Eq.(\ref{eq:flux}). The amplitude of $\Delta l>1$ transition is nonzero for twisted photons with $m_\gamma=m_f$, as shown in 
Eq.(\ref{eq:lowb-anyl}), implying that nonzero transition rates take place at the node of the field. The result is not unphysical, because the transition amplitude calculated in Eqs.(\ref{eq:genresult},\ref{eq:reduced}) is essentially non-local quantity since it involves infinite radial integrals over the wave function with a characteristic size of $a_0$, and the regions with nonzero field contribute to the result.

\


\section{Summary and Discussion}			\label{sec:disc}

It this work we applied previously developed theoretical formalism \cite{Afanasev:2013kaa,Afanasev2014,Scholz2014} to analyze photoexcitation of an atom by Bessel beams with high angular momentum.
We demonstrate that the excitation rates are finite near the zero-intensity region of an optical vortex center for $l_f\ge m_\gamma>1$, i.e. for any final states with large angular momentum that match the twisted-photon quantum numbers. We also show that the rates of  $\Delta l>1$ transitions are enhanced relative to the plane-wave photons, with the most dramatic effect observable near the vortex center.
However, for twisted photons $\Delta l>1$ transition rates remain much smaller than $\Delta l=1$ electric-dipole rates, unless the atom is placed near the vortex center at atomic-scale distances. Since attenuation of light in the atomic matter is controlled by the largest $E1$-transitions whose rates are proportional to the beam intensity across the wavefront, the light would be attenuated by the same percentage independently of proximity to the optical vortex center for such transitions. However, our results have surprising implications for higher multipoles, namely, they indicate that photon's attenuation in homogeneous atomic matter should depend on the proximity to the optical vortex center, where it reaches the maximum.
We obtained compact expressions for quantum selection rules near the vortex center that allow simple estimates of the strengths of multipole transitions caused by the twisted waves.

Awhile  ago Berry and Dennis introduced a "quantum core" concept \cite{BerryDennis2004}, according to which the nodal line singularities (optical vortices) are smoothed in quantum optics because of spontaneous emission into unoccupied modes. In the present work we demonstrate that when using an atom as a probe of the optical vortex, the node of the electromagnetic field in the vortex center is smoothed by higher-multipole ($\Delta l>1$) atomic excitations. We find that the characteristic size of this "quantum core"  is Bohr radius, or a size of an atomic probe. Therefore experimental observation of the "excitation in the dark" by an optical vortex \cite{Schmiegelow2015} may be considered as a "quantum core" effect. Here we emphasize that "excitation in the dark" takes place for any higher multipoles with $l_f\ge m_\gamma>1$. While the quadrupole E2 transitions near phase singularity are driven by the field {\it gradients}, the second-order field derivative (or {\it field curvature}) is responsible for the $E3$ octupole transitions.

In summary, we pointed out the unique features that make the interaction of the twisted light with atomic matter different from the plane waves. We believe these results will be helpful in de-coding quantum-level information that can be transmitted by the twisted light and detected by well-localized probes such as atoms, molecules or nano-structures. 


\begin{acknowledgments}

CEC thanks the National Science Foundation for support under Grant
PHY-1516509. Work of AA was supported by Gus Weiss Foundation of The George  Washington University. 
Useful discussions with K. Bliokh, S. Franke-Arnold, M.V.~Berry and M.R.~Dennis are gratefully acknowledged.

\end{acknowledgments}



\bibliography{TwistedPhoton}

\begin{thebibliography}{24}
\expandafter\ifx\csname natexlab\endcsname\relax\def\natexlab#1{#1}\fi
\expandafter\ifx\csname bibnamefont\endcsname\relax
  \def\bibnamefont#1{#1}\fi
\expandafter\ifx\csname bibfnamefont\endcsname\relax
  \def\bibfnamefont#1{#1}\fi
\expandafter\ifx\csname citenamefont\endcsname\relax
  \def\citenamefont#1{#1}\fi
\expandafter\ifx\csname url\endcsname\relax
  \def\url#1{\texttt{#1}}\fi
\expandafter\ifx\csname urlprefix\endcsname\relax\def\urlprefix{URL }\fi
\providecommand{\bibinfo}[2]{#2}
\providecommand{\eprint}[2][]{\url{#2}}

\bibitem[{\citenamefont{Allen et~al.}(1992)\citenamefont{Allen, Beijersbergen,
  Spreeuw, and Woerdman}}]{Allen:1992zz}
\bibinfo{author}{\bibfnamefont{L.}~\bibnamefont{Allen}},
  \bibinfo{author}{\bibfnamefont{M.}~\bibnamefont{Beijersbergen}},
  \bibinfo{author}{\bibfnamefont{R.}~\bibnamefont{Spreeuw}}, \bibnamefont{and}
  \bibinfo{author}{\bibfnamefont{J.}~\bibnamefont{Woerdman}},
  \bibinfo{journal}{Phys.Rev.} \textbf{\bibinfo{volume}{A45}},
  \bibinfo{pages}{8185} (\bibinfo{year}{1992}).

\bibitem[{\citenamefont{Molina-Terriza
  et~al.}(2007)\citenamefont{Molina-Terriza, Torres, and
  Torner}}]{molina2007nature}
\bibinfo{author}{\bibfnamefont{G.}~\bibnamefont{Molina-Terriza}},
  \bibinfo{author}{\bibfnamefont{J.~P.} \bibnamefont{Torres}},
  \bibnamefont{and} \bibinfo{author}{\bibfnamefont{L.}~\bibnamefont{Torner}},
  \bibinfo{journal}{Nature Physics} \textbf{\bibinfo{volume}{3}},
  \bibinfo{pages}{305} (\bibinfo{year}{2007}).

\bibitem[{\citenamefont{Yao and Padgett}(2011)}]{Yao11}
\bibinfo{author}{\bibfnamefont{A.}~\bibnamefont{Yao}} \bibnamefont{and}
  \bibinfo{author}{\bibfnamefont{M.}~\bibnamefont{Padgett}},
  \bibinfo{journal}{Advances in Optics and Photonics}
  \textbf{\bibinfo{volume}{3}}, \bibinfo{pages}{161} (\bibinfo{year}{2011}).

\bibitem[{\citenamefont{Wisniewski-€Barker and Padgett}(2015)}]{Padgett2015}
\bibinfo{author}{\bibfnamefont{E.}~\bibnamefont{Wisniewski-€Barker}}
  \bibnamefont{and} \bibinfo{author}{\bibfnamefont{M.}~\bibnamefont{Padgett}},
  \bibinfo{journal}{Photonics: Scientific Foundations, Technology and
  Applications,} \textbf{\bibinfo{volume}{1}}, \bibinfo{pages}{321}
  (\bibinfo{year}{2015}).

\bibitem[{\citenamefont{Afanasev et~al.}(2013)\citenamefont{Afanasev, Carlson,
  and Mukherjee}}]{Afanasev:2013kaa}
\bibinfo{author}{\bibfnamefont{A.}~\bibnamefont{Afanasev}},
  \bibinfo{author}{\bibfnamefont{C.~E.} \bibnamefont{Carlson}},
  \bibnamefont{and}
  \bibinfo{author}{\bibfnamefont{A.}~\bibnamefont{Mukherjee}},
  \bibinfo{journal}{Phys. Rev. A,} \textbf{\bibinfo{volume}{88}},
  \bibinfo{pages}{033841} (\bibinfo{year}{2013}).

\bibitem[{\citenamefont{Afanasev et~al.}(2014)\citenamefont{Afanasev, Carlson,
  and Mukherjee}}]{Afanasev2014}
\bibinfo{author}{\bibfnamefont{A.}~\bibnamefont{Afanasev}},
  \bibinfo{author}{\bibfnamefont{C.~E.} \bibnamefont{Carlson}},
  \bibnamefont{and}
  \bibinfo{author}{\bibfnamefont{A.}~\bibnamefont{Mukherjee}},
  \bibinfo{journal}{J. Opt. Soc. Am. B} \textbf{\bibinfo{volume}{31}},
  \bibinfo{pages}{2721} (\bibinfo{year}{2014}).

\bibitem[{\citenamefont{Scholz-Marggraf
  et~al.}(2014)\citenamefont{Scholz-Marggraf, Fritzsche, Serbo, Afanasev, and
  Surzhykov}}]{Scholz2014}
\bibinfo{author}{\bibfnamefont{H.~M.} \bibnamefont{Scholz-Marggraf}},
  \bibinfo{author}{\bibfnamefont{S.}~\bibnamefont{Fritzsche}},
  \bibinfo{author}{\bibfnamefont{V.~G.} \bibnamefont{Serbo}},
  \bibinfo{author}{\bibfnamefont{A.}~\bibnamefont{Afanasev}}, \bibnamefont{and}
  \bibinfo{author}{\bibfnamefont{A.}~\bibnamefont{Surzhykov}},
  \bibinfo{journal}{Phys. Rev. A} \textbf{\bibinfo{volume}{90}},
  \bibinfo{pages}{013425} (\bibinfo{year}{2014}),
  \urlprefix\url{http://link.aps.org/doi/10.1103/PhysRevA.90.013425}.

\bibitem[{\citenamefont{Surzhykov et~al.}(2015)\citenamefont{Surzhykov, Seipt,
  Serbo, and Fritzsche}}]{Surzhikov15}
\bibinfo{author}{\bibfnamefont{A.}~\bibnamefont{Surzhykov}},
  \bibinfo{author}{\bibfnamefont{D.}~\bibnamefont{Seipt}},
  \bibinfo{author}{\bibfnamefont{V.~G.} \bibnamefont{Serbo}}, \bibnamefont{and}
  \bibinfo{author}{\bibfnamefont{S.}~\bibnamefont{Fritzsche}},
  \bibinfo{journal}{Phys. Rev. A} \textbf{\bibinfo{volume}{91}},
  \bibinfo{pages}{013403} (\bibinfo{year}{2015}),
  \urlprefix\url{http://link.aps.org/doi/10.1103/PhysRevA.91.013403}.

\bibitem[{\citenamefont{{Rodrigues} et~al.}(2015)\citenamefont{{Rodrigues},
  {Marcassa}, and {Mendon{\c c}a}}}]{Rodrigues2015}
\bibinfo{author}{\bibfnamefont{J.~D.} \bibnamefont{{Rodrigues}}},
  \bibinfo{author}{\bibfnamefont{L.~G.} \bibnamefont{{Marcassa}}},
  \bibnamefont{and} \bibinfo{author}{\bibfnamefont{J.~T.}
  \bibnamefont{{Mendon{\c c}a}}}, \bibinfo{journal}{ArXiv e-prints}
  (\bibinfo{year}{2015}), \eprint{1512.05933}.

\bibitem[{\citenamefont{J\'auregui}(2015)}]{Jaregui2015}
\bibinfo{author}{\bibfnamefont{R.}~\bibnamefont{J\'auregui}},
  \bibinfo{journal}{Phys. Rev. A} \textbf{\bibinfo{volume}{91}},
  \bibinfo{pages}{043842} (\bibinfo{year}{2015}),
  \urlprefix\url{http://link.aps.org/doi/10.1103/PhysRevA.91.043842}.

\bibitem[{\citenamefont{Mondal et~al.}(2014)\citenamefont{Mondal, Deb, and
  Majumder}}]{Mondal2014}
\bibinfo{author}{\bibfnamefont{P.~K.} \bibnamefont{Mondal}},
  \bibinfo{author}{\bibfnamefont{B.}~\bibnamefont{Deb}}, \bibnamefont{and}
  \bibinfo{author}{\bibfnamefont{S.}~\bibnamefont{Majumder}},
  \bibinfo{journal}{Phys. Rev. A} \textbf{\bibinfo{volume}{89}},
  \bibinfo{pages}{063418} (\bibinfo{year}{2014}),
  \urlprefix\url{http://link.aps.org/doi/10.1103/PhysRevA.89.063418}.

\bibitem[{\citenamefont{Kaplan and McGuire}(2015)}]{Kaplan15}
\bibinfo{author}{\bibfnamefont{L.}~\bibnamefont{Kaplan}} \bibnamefont{and}
  \bibinfo{author}{\bibfnamefont{J.~H.} \bibnamefont{McGuire}},
  \bibinfo{journal}{Phys. Rev. A} \textbf{\bibinfo{volume}{92}},
  \bibinfo{pages}{032702} (\bibinfo{year}{2015}),
  \urlprefix\url{http://link.aps.org/doi/10.1103/PhysRevA.92.032702}.

\bibitem[{\citenamefont{Babiker et~al.}(2002)\citenamefont{Babiker, Bennett,
  Andrews, and D\'avila~Romero}}]{Babiker2002}
\bibinfo{author}{\bibfnamefont{M.}~\bibnamefont{Babiker}},
  \bibinfo{author}{\bibfnamefont{C.~R.} \bibnamefont{Bennett}},
  \bibinfo{author}{\bibfnamefont{D.~L.} \bibnamefont{Andrews}},
  \bibnamefont{and} \bibinfo{author}{\bibfnamefont{L.~C.}
  \bibnamefont{D\'avila~Romero}}, \bibinfo{journal}{Phys. Rev. Lett.}
  \textbf{\bibinfo{volume}{89}}, \bibinfo{pages}{143601}
  (\bibinfo{year}{2002}),
  \urlprefix\url{http://link.aps.org/doi/10.1103/PhysRevLett.89.143601}.

\bibitem[{\citenamefont{Pic\'on et~al.}(2010)\citenamefont{Pic\'on, Mompart,
  de~Aldana, Plaja, Calvo, and Roso}}]{Picon10}
\bibinfo{author}{\bibfnamefont{A.}~\bibnamefont{Pic\'on}},
  \bibinfo{author}{\bibfnamefont{J.}~\bibnamefont{Mompart}},
  \bibinfo{author}{\bibfnamefont{J.~R.~V.} \bibnamefont{de~Aldana}},
  \bibinfo{author}{\bibfnamefont{L.}~\bibnamefont{Plaja}},
  \bibinfo{author}{\bibfnamefont{G.~F.} \bibnamefont{Calvo}}, \bibnamefont{and}
  \bibinfo{author}{\bibfnamefont{L.}~\bibnamefont{Roso}},
  \bibinfo{journal}{Optics Express} \textbf{\bibinfo{volume}{18}},
  \bibinfo{pages}{3660} (\bibinfo{year}{2010}), \eprint{1002.1318}.

\bibitem[{\citenamefont{{Schmiegelow} et~al.}(2015)\citenamefont{{Schmiegelow},
  {Schulz}, {Kaufmann}, {Ruster}, {Poschinger}, and
  {Schmidt-Kaler}}}]{Schmiegelow2015}
\bibinfo{author}{\bibfnamefont{C.~T.} \bibnamefont{{Schmiegelow}}},
  \bibinfo{author}{\bibfnamefont{J.}~\bibnamefont{{Schulz}}},
  \bibinfo{author}{\bibfnamefont{H.}~\bibnamefont{{Kaufmann}}},
  \bibinfo{author}{\bibfnamefont{T.}~\bibnamefont{{Ruster}}},
  \bibinfo{author}{\bibfnamefont{U.~G.} \bibnamefont{{Poschinger}}},
  \bibnamefont{and}
  \bibinfo{author}{\bibfnamefont{F.}~\bibnamefont{{Schmidt-Kaler}}},
  \bibinfo{journal}{ArXiv e-prints}  (\bibinfo{year}{2015}),
  \eprint{1511.07206}.

\bibitem[{\citenamefont{Schmiegelow and Schmidt-Kaler}(2012)}]{Schmiegelow2012}
\bibinfo{author}{\bibfnamefont{C.}~\bibnamefont{Schmiegelow}} \bibnamefont{and}
  \bibinfo{author}{\bibfnamefont{F.}~\bibnamefont{Schmidt-Kaler}},
  \bibinfo{journal}{Eur. Phys. J. D} \textbf{\bibinfo{volume}{66}},
  \bibinfo{pages}{157} (\bibinfo{year}{2012}),
  \urlprefix\url{http://dx.doi.org/10.1140/epjd/e2012-20730-4}.

\bibitem[{\citenamefont{Jentschura and
  Serbo}(2011{\natexlab{a}})}]{Jentschura:2010ap}
\bibinfo{author}{\bibfnamefont{U.}~\bibnamefont{Jentschura}} \bibnamefont{and}
  \bibinfo{author}{\bibfnamefont{V.}~\bibnamefont{Serbo}},
  \bibinfo{journal}{Phys.Rev.Lett.} \textbf{\bibinfo{volume}{103}},
  \bibinfo{pages}{013001} (\bibinfo{year}{2011}{\natexlab{a}}), \eprint{eprint
  arXiv:1008.4788}.

\bibitem[{\citenamefont{Jentschura and
  Serbo}(2011{\natexlab{b}})}]{Jentschura:2011ih}
\bibinfo{author}{\bibfnamefont{U.}~\bibnamefont{Jentschura}} \bibnamefont{and}
  \bibinfo{author}{\bibfnamefont{V.}~\bibnamefont{Serbo}},
  \bibinfo{journal}{Eur.Phys.J.} \textbf{\bibinfo{volume}{C71}},
  \bibinfo{pages}{1571} (\bibinfo{year}{2011}{\natexlab{b}}), \eprint{eprint
  arXiv: 1101.1206}.

\bibitem[{\citenamefont{Durnin}(1987)}]{Durnin:1987}
\bibinfo{author}{\bibfnamefont{J.}~\bibnamefont{Durnin}}, \bibinfo{journal}{J.
  Opt. Soc. Am. A} \textbf{\bibinfo{volume}{4}}, \bibinfo{pages}{651}
  (\bibinfo{year}{1987}).

\bibitem[{\citenamefont{Durnin et~al.}(1987)\citenamefont{Durnin, Miceli, and
  Eberly}}]{Durnin:1987zz}
\bibinfo{author}{\bibfnamefont{J.}~\bibnamefont{Durnin}},
  \bibinfo{author}{\bibfnamefont{J.}~\bibnamefont{Miceli}}, \bibnamefont{and}
  \bibinfo{author}{\bibfnamefont{J.}~\bibnamefont{Eberly}},
  \bibinfo{journal}{Phys. Rev. Lett.} \textbf{\bibinfo{volume}{58}},
  \bibinfo{pages}{1499} (\bibinfo{year}{1987}).

\bibitem[{\citenamefont{Jauregui}(2004)}]{Jauregui:2004}
\bibinfo{author}{\bibfnamefont{R.}~\bibnamefont{Jauregui}},
  \bibinfo{journal}{Phys.Rev.} \textbf{\bibinfo{volume}{A70}},
  \bibinfo{pages}{033415} (\bibinfo{year}{2004}).

\bibitem[{\citenamefont{Bliokh et~al.}(2010)\citenamefont{Bliokh, Alonso,
  Ostrovskaya, and Aiello}}]{BliokhAlonso10}
\bibinfo{author}{\bibfnamefont{K.~Y.} \bibnamefont{Bliokh}},
  \bibinfo{author}{\bibfnamefont{M.~A.} \bibnamefont{Alonso}},
  \bibinfo{author}{\bibfnamefont{E.~A.} \bibnamefont{Ostrovskaya}},
  \bibnamefont{and} \bibinfo{author}{\bibfnamefont{A.}~\bibnamefont{Aiello}},
  \bibinfo{journal}{Phys. Rev. A} \textbf{\bibinfo{volume}{82}},
  \bibinfo{pages}{063825} (\bibinfo{year}{2010}),
  \urlprefix\url{http://link.aps.org/doi/10.1103/PhysRevA.82.063825}.

\bibitem[{\citenamefont{Bliokh and Nori}(2015)}]{Bliokh2015}
\bibinfo{author}{\bibfnamefont{K.~Y.} \bibnamefont{Bliokh}} \bibnamefont{and}
  \bibinfo{author}{\bibfnamefont{F.}~\bibnamefont{Nori}},
  \bibinfo{journal}{Physics Reports} \textbf{\bibinfo{volume}{592}},
  \bibinfo{pages}{1 } (\bibinfo{year}{2015}), ISSN \bibinfo{issn}{0370-1573},
  \bibinfo{note}{transverse and longitudinal angular momenta of light},
  \urlprefix\url{http://www.sciencedirect.com/science/article/pii/S0370157315003336}.

\bibitem[{\citenamefont{Berry and Dennis}(2004)}]{BerryDennis2004}
\bibinfo{author}{\bibfnamefont{M.~V.} \bibnamefont{Berry}} \bibnamefont{and}
  \bibinfo{author}{\bibfnamefont{M.~R.} \bibnamefont{Dennis}},
  \bibinfo{journal}{Journal of Optics A: Pure and Applied Optics}
  \textbf{\bibinfo{volume}{6}}, \bibinfo{pages}{S178} (\bibinfo{year}{2004}),
  \urlprefix\url{http://stacks.iop.org/1464-4258/6/i=5/a=006}.

\end{thebibliography}

\end{document}